\documentclass[
 reprint,
 superscriptaddress,
 amsmath,amssymb,
 aps,
 pra, 
]{revtex4-2}

\usepackage{graphicx}
\usepackage{dcolumn}
\usepackage{multirow} 
\usepackage{xcolor}
\usepackage[normalem]{ulem}
\usepackage{bm}
\usepackage{hyperref}
\usepackage{physics}
\usepackage{braket}
\usepackage{amsmath}
\usepackage{amsfonts}
\usepackage{amssymb}
\usepackage{amsthm}
\usepackage{mathtools}
\usepackage{color} 
\usepackage{gensymb}
\usepackage{here}

\begin{document}

\title{Supramolecular approach-based intermolecular interaction energy calculations using quantum phase estimation algorithm}

\author{Yuhei Tachi}
\email{yuhei.tachi@tohmatsu.co.jp}
\affiliation{Deloitte Tohmatsu LLC, 3-2-3 Marunouchi, Chiyoda-ku, Tokyo 100-8363, Japan}

\author{Akihiko Arakawa}
\affiliation{Chugai Pharmaceutical Co., Ltd., 216 Totsuka-cho, Totsuka-ku, Yokohama, Kanagawa 244-0003, Japan}

\author{Taisei Osawa}
\affiliation{Chugai Pharmaceutical Co., Ltd., 216 Totsuka-cho, Totsuka-ku, Yokohama, Kanagawa 244-0003, Japan}

\author{Masayoshi Terabe}
\affiliation{Deloitte Tohmatsu LLC, 3-2-3 Marunouchi, Chiyoda-ku, Tokyo 100-8363, Japan}

\author{Kenji Sugisaki}
\email{kensugisaki@tohmatsu.co.jp}
\affiliation{Deloitte Tohmatsu LLC, 3-2-3 Marunouchi, Chiyoda-ku, Tokyo 100-8363, Japan}
\affiliation{Centre for Quantum Engineering, Research and Education (CQuERE), TCG Centres for Research and Education in Science and Technology (TCG CREST), Sector V, Salt Lake, Kolkata 700091, India}

\date{\today}

\begin{abstract}
Accurate computation of non-covalent, intermolecular interaction energies is important to understand various chemical phenomena, and quantum computers are anticipated to accelerate it. Although the state-of-the-art quantum computers are still noisy and intermediate-scale ones, development of theoretical frameworks those are expected to work on a fault-tolerant quantum computer is an urgent issue. In this work, we explore resource-efficient implementation of the quantum phase estimation-based complete active space configuration interaction (QPE-CASCI) calculations, with the aid of the second-order M{\o}ller--Plesset perturbation theory (MP2)-based active space selection with Boys localized orbitals. We performed numerical simulations of QPE for the supramolecular approach-based intermolecular interaction energy calculations of the hydrogen-bonded water dimer, using 6 system and 6 ancilla qubits. With the aid of algorithmic error mitigation, the QPE-CASCI simulations achieved interaction energy predictions with an error of 0.02 kcal mol$^{-1}$ relative to the CASCI result, demonstrating the accuracy and efficiency of the proposed methodology. Preliminary results on quantum circuit compression for QPE are also presented to reduce the number of two-qubit gates and
depth. 
\end{abstract}
\maketitle

\section{Introduction}
Non-covalent, intermolecular interactions play a pivotal role in a wide range of chemical and physical phenomena, influencing molecular recognition, self-assembly, and reactivity. These interactions, which include hydrogen bonding, van der Waals forces, $\pi$--$\pi$ stacking, and electrostatic interactions, are central to the behavior of complex systems in chemistry, biology, and materials science~\cite{Puzzarini-2020}. For example, intermolecular interaction between protein and pharmaceutical molecules is one of the key factors that control efficacy of drugs~\cite{Bissantz-2010, Bissantz-2010_correction}.
Traditional computational methods such as density functional theory (DFT) have been widely used to calculate intermolecular interaction energies, but DFT often face limitations when handling strongly correlated systems such as multinuclear transition metal complexes appear as active sites of enzymes, or achieving high precision for energy estimation~\cite{Sousa-2007, Cohen-2012}.

Quantum computing has emerged as one of the most promising technologies in current science, and a large number of studies on quantum computation have been reported from both experimental and theoretical perspectives. To date, quantum algorithms that can improve the computational cost scaling against classical counterpart have been reported; for example, prime number factorization~\cite{Shor-1994}, solving linear system of equations~\cite{Harrow-2009}, and finding eigenvalue of a unitary operator~\cite{Abrams-1999}. Among diverse topics in quantum computing, solving electronic structures of atoms and molecules, namely quantum chemical calculations, has been anticipated as important applications of quantum computers~\cite{Cao-2019, Bauer-2020}. In the {\it ab initio} electronic structure theory, the computational cost of the full configuration interaction (full-CI) calculation, which gives variationally best wave function within the Hilbert space spanned by the basis set used, scales exponentially with the system size, while quantum computer is capable of computing the full-CI energy in polynomial time using quantum phase estimation (QPE) algorithm~\cite{Abrams-1999, Aspuru-2005}, if the approximate wave function having sufficiently large overlap with the exact wave function is accessible with a polynomial cost. It should be noted that although the QPE itself belongs to computational complexity class {\bf BQP} and it can be executed on a quantum computer in polynomial time~\cite{Wocjan-2006}, preparation of sophisticated approximate wave function is in general very difficult, especially for large molecules like biomolecules and strongly correlated systems~\cite{Lee-2023}. In fact, preparation of an eigenfunction of $k$-local Hamiltonian with $k \ge 2$ belongs to the complexity class {\bf QMA}~\cite{Kempe-2006}. 

There are two famous approaches to investigate intermolecular interaction energy using quantum chemical calculations, namely supramolecular approach and symmetry-adapted perturbation theory (SAPT)~\cite{Szalewicz-2011}. In the supramolecular approach, an intermolecular interaction energy $E_\text{int}$ is calculated using the following equation, 
\begin{eqnarray}
    E_\text{int} = E_\text{AB} - E_\text{A} - E_\text{B}, 
    \label{eq1}
\end{eqnarray}
where $E_\text{A}$, $E_\text{B}$, and $E_\text{AB}$ are energies of monomer A monomer B, and interacting system (A$\cdots$B), respectively. Application of quantum computation to supramolecular approach is straightforward, because there are known quantum computational methods for total energies, such as QPE~\cite{Aspuru-2005}, quantum-selected configuration interaction (QSCI)~\cite{Kanno-2023, Robledo-2025}, and variational quantum eigensolver (VQE)~\cite{Peruzzo-2014, Tilly-2022}. Recently, non-covalent interactions in water and methane dimers have been investigated using superconducting quantum devices with the QSCI method in conjunction with self-consistent configuration recovery~\cite{Kaliakin-2024}. It is important to note that the QSCI method does not inherently satisfy the size consistency condition~\cite{Szabo-Ostlund} without special care. Therefore, the intermolecular interaction energy was calculated as the difference between the total energies of the interacting and non-interacting dimers.
In SAPT, an intermolecular interaction energy is calculated based on perturbation theory, using the wave functions of monomers as the unperturbed wave functions. Note that quantum computational approaches to SAPT have been proposed both in noisy intermediate-scale quantum (NISQ)~\cite{Loipersberger-2023, Ollitrault-2024} and in fault-tolerant quantum computer (FTQC)~\cite{Cortes-2024} frameworks. 
In this work, we numerically demonstrate intermolecular interaction energy calculations using QPE algorithm~\cite{Abrams-1999, Aspuru-2005} in conjunction with the supramolecular approach. To this end, we focus on typical and essentially important intermolecular interaction, namely hydrogen bond between water molecules, as an example. We also calculate potential energy curve of water dimer by changing intermolecular distance. To develop a resource-efficient computational workflow, we adopted second-order M{\o}ller--Plesset perturbation theory (MP2)-based active space selection with Boys localized orbitals~\cite{Boys-1960}. 

This paper is organized as follows. In Section II, we provides theoretical backgrounds of the QPE algorithm and orbital localizations. Detailed computational conditions are given in Section III. The results of quantum chemical calculations and quantum circuit simulations are given in Section IV. Conclusions and future perspectives are provided in Section V.  

\section{Theory}
\subsection{Quantum phase estimation}
In the field of quantum computation of molecular electronic structures, QPE is well known approach to calculate the full-CI energy, or complete active space configuration interaction (CASCI) energy when the active space approximation is introduced. QPE relies on the fact that the time evolution of an eigenfunction of Hamiltonian $H$ causes a phase shift $\phi$ depending on the corresponding eigenvalue $E$, 
\begin{eqnarray}
    e^{-iCHt}|\Psi\rangle = e^{-iCEt}|\Psi\rangle = e^{i2\pi\phi}|\Psi\rangle.
    \label{eq2}
\end{eqnarray}
Here, a scaling factor $C$ is introduced to satisfy $0 \le \phi < 1$. The inverse of the $L1$ norm of Hamiltonian $1/||H||$ is often used as a scaling factor $C$. However, this scaling factor is usually too conservative for quantum chemical calculations, because the Hartree--Fock (HF) energy $E_\text{HF}$ provides an upper bound of the full-CI energy, and the HF usually covers about 99\% of the total energy. In this work, we calculate the correlation energy $E_\text{Corr} = E - E_\text{HF}$ instead of the total energy $E$ using QPE, by replacing Hamiltonian $H$ by $H - E_\text{HF}$, as follows. 
\begin{eqnarray}
    e^{-iC(H-E_\text{HF})t}|\Psi\rangle = e^{-iCE_\text{Corr}t}|\Psi\rangle = e^{i2\pi\phi}|\Psi\rangle
    \label{eq3}
\end{eqnarray}
In this case, the appropriate evolution time length with a scaling factor $Ct$ can be conveniently determined, based on the electron correlation energy computed at the MP2 level ($E_\text{Corr}^{(2)}$). For example, we can set as $Ct = 5\times |E_\text{Corr}^{(2)}|$. In the following discussion, we use $C = 1$ and omit $C$ in the equation. 

\begin{figure}
    \centering
    \includegraphics[width=\linewidth]{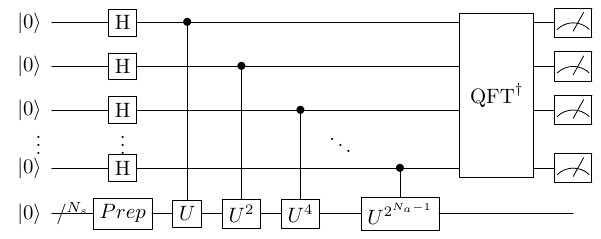}
    \caption{Quantum circuit for QPE.}
    \label{fig:fig1}
\end{figure}

The quantum circuit of QPE used in this work is shown in Figure \ref{fig:fig1}. Here, H is a Hadamard gate and $U = e^{-i(H - E_\text{HF})t}$. The QPE quantum circuit consists of $N_s$ of system qubits to store wave function and $N_a$ of ancillary qubits to read out the phase. The quantum circuit is composed of three steps: State preparation, controlled-time evolution operation, and inverse quantum Fourier transform. The state preparation gate acting on the system qubits generates an approximate wave function $|\Phi\rangle$ of the target electronic state, 
\begin{eqnarray}
    Prep|0\rangle^{\otimes N_s} = |\Phi\rangle = \sum_j c_j |\Psi_j\rangle. 
    \label{eq4}
\end{eqnarray}
Here, $|\Psi_j\rangle$ represents $j$-th eigenfunction of Hamiltonian, and $c_j$ is a corresponding expansion coefficient. %We assume that the approximate wave function has polynomially large overlap with the target eigenstate, $|\langle \Phi | \Psi_{target}\rangle| \approx Poly(N_s)$. 
After applying Hadamard gates to the ancillary qubits, controlled-time evolutions, and subsequent inverse quantum Fourier transform, the quantum state before the measurement is given as follows, 
\begin{eqnarray}
    \sum_j c'_j |\phi'_j\rangle \otimes |\Psi_j\rangle, 
    \label{eq5}
\end{eqnarray}
and measurement of ancillary qubits projects the quantum state to one of the eigenstate with a probability proportional to $|c'_j|^2$, and the measured bitstrings $x_1 x_2 \cdots x_{N_a}$ corresponds to the eigenphase in binary fraction, $\phi' = \sum_{k=1}^{N_a} x_k/2^k$. The projective nature of the QPE algorithm allows us to compute eigenvalue of Hamiltonian even if approximate wave function is used. When the system is weakly correlated, as in the case of water molecules under study, the HF wave function is a reasonable choice of the input wave function in QPE.  

Note that when the true eigenphase $\phi$ has a $N_a$-bit expansion, it holds that $\phi' = \phi$ and $|c'_j|^2 = |c_j|^2$. However, this condition is typically not satisfied, resulting in multiple phase values close to $\phi$ being obtained in QPE, which is known as a spectral leakage. Note that the spectral leakage can be mitigated by using appropriate entangling gates instead of Hadamard gates for state preparation of ancillary qubits~\cite{Sakuma-2025}. 

The most cost demanding part of the QPE quantum circuit is controlled-time evolution operations. To implement the time evolution operation on a quantum computer, the second-quantized Hamiltonian given as
\begin{eqnarray}
    H = \sum_{pq} h_{pq} a_p^\dagger a_q + \frac{1}{2}\sum_{pqrs} g_{pqrs} a_p^\dagger a_q^\dagger a_s a_r
    \label{eq6}
\end{eqnarray}
is transformed to a qubit Hamiltonian
\begin{eqnarray}
    H_\text{qub} = \sum_l^L \omega_l P_l, 
    \label{eq7}
\end{eqnarray}
where $P_l$ is a direct product of Pauli operators termed as a Pauli string, and $\omega_l$ is a corresponding coefficient computed from $h_{pq}$ and $g_{pqrs}$. Several fermion--qubit encoding techniques have been reported, such as Jordan--Wigner transformation (JWT)~\cite{Jordan-1928}, Bravyi--Kitaev transformation (BKT)~\cite{Seeley-2012}, and symmetry-conserving Bravyi--Kitaev transformation (SCBKT)~\cite{Bravyi-2017}. Because qubit Hamiltonian $H_\text{qub}$ generally contains operators those are not commute each other, Trotter--Suzuki decomposition is often applied to construct the quantum circuit for the time evolution simulations. For example, the second-order Trotter--Suzuki decomposition is given as
\begin{eqnarray}
    e^{-i H_\text{qub} t} = \left[\Pi_{l=1}^L e^{-i \omega_l P_l t/2M} \Pi_{l=L}^1 e^{-i\omega_l p_l t/2M} \right]^M,
    \label{eq8}
\end{eqnarray}
and the quantum circuit for the Trotterized time evolution operator is constructed by sequentially arranging the quantum circuit corresponding to $e^{-i \omega_l P_l t/2M}$ operation. It should be noted that the error of Trotter decomposition depends not only on the order of the Trotterization but also on the sequence of the Trotterized terms, and adopting appropriate operator ordering is important to obtain accurate QPE energy~\cite{Tranter-2018, Tranter-2019}.
Note that QPE implementations based on other approximate time-propagation methods have also been reported~\cite{Berry-2015, Low-2017, Low-2019, Martyn-2021}. 

In principle, the time evolution operator $e^{-iHt}$ and Hamiltonian $H$ commutes each other, but Trotterized time evolution operator does not necessary to commute with Hamiltonian. As a result, Trotter decomposition error can potentially lead to the loss of important properties inherent to full-CI. In fact, recent numerical studies revealed that Trotter decomposition of time evolution operator can break size consistency condition~\cite{Szabo-Ostlund} in QPE~\cite{Sugisaki-2024b}. Note that breakdown of size consistency caused by Trotter decomposition has also been observed in VQE with unitary coupled cluster singles and doubles (UCCSD) ansatz~\cite{Sugisaki-2024a}. The study also disclosed that using molecular orbitals localized to monomer and adopting magnitude ordering~\cite{Tranter-2018} for the Trotterized term ordering efficiently suppress the violation of size consistency~\cite{Sugisaki-2024b}. Because size consistency is essential for the interaction energy calculation based on supramolecular approach, we use localized molecular orbitals for the QPE numerical simulations. Orbital localization is also useful to reduce computational cost by excluding molecular orbitals those are less important to the interaction energy calculation from the active space, as discussed below. 

Note that the QPE with active space approximation returns the CASCI energy and therefore large part of dynamical electron correlation effects are not taken into account. Connecting other quantum algorithms for dynamical electron correlation effects such as the MP2~\cite{Mitarai-2023} is essential to obtain the quantitative energy, which is out of scope of this study.  

\subsection{Molecular orbital localization and MP2 correlation energy-based active orbital selection}
As we pointed out in the previous section, using molecular orbitals localized to monomer is a key ingredient to maintain size consistency condition in QPE with Trotterization. In addition, in order to reduce computational cost by excluding molecular orbitals those are unimportant to the intermolecular interactions, adopting the orbital localization method that gives approximately the same localized orbitals for the monomer and the dimer is required. There are several known orbital localization techniques depending on the definition of the operator. For example, Boys localization scheme uses the square of the distance between two electrons as the operator, and minimizes the expectation value~\cite{Boys-1960}. In the Edmiston--Ruedenberg (ER) localization scheme, an expectation value of the inverse of the distance between two electrons is maximized~\cite{Edmiston-1963, Edmiston-1965}. Pipek--Mezey (PM) maximizes the sum of the Mulliken atomic charges~\cite{Pipek-1989}. As we discuss in Results and Discussion section, we examined these three localization methods, and adopted Boys localization for the QPE simulations. 

Since localized orbitals are spatially confined to a relatively small volume, we expect that we can use smaller active space for the interaction energy calculations than in the HF canonical orbital-based calculations, by excluding the molecular orbitals spatially well separated from the interaction site. One approach to select active orbitals is based on their spatial distribution, but we found that applying this strategy is sometimes difficult. Instead, we investigated the active orbital selection strategy based on the MP2 correlation energies.  

The MP2 correlation energy can be calculated using the following equation~\cite{Szabo-Ostlund}, 
\begin{eqnarray}
    E_{\rm Corr}^{(2)} = \frac{1}{4} \sum_{ijab} \frac{|g_{ijba} - g_{ijab}|^2}{\varepsilon_i + \varepsilon_j - \varepsilon_a - \varepsilon_b},
    \label{eq9}
\end{eqnarray}
where $i$ and $j$ represent occupied molecular orbitals and $a$ and $b$ stand for unoccupied molecular orbitals. $\varepsilon_i$ is the orbital energy of the $i$-th molecular orbital. Note that orbital energy is not uniquely defined in the localized orbitals, as they are not eigenfunctions of a Fock operator. In the present study, we used expectation values of the Fock operator as alternatives of the orbital energies. As a metric for active orbital selection, we computed orbital-wise and excitation-wise correlation energy contributions, by taking partial sum of the correlation energy.  

\begin{figure*}
    \centering
    \includegraphics[scale=0.6]{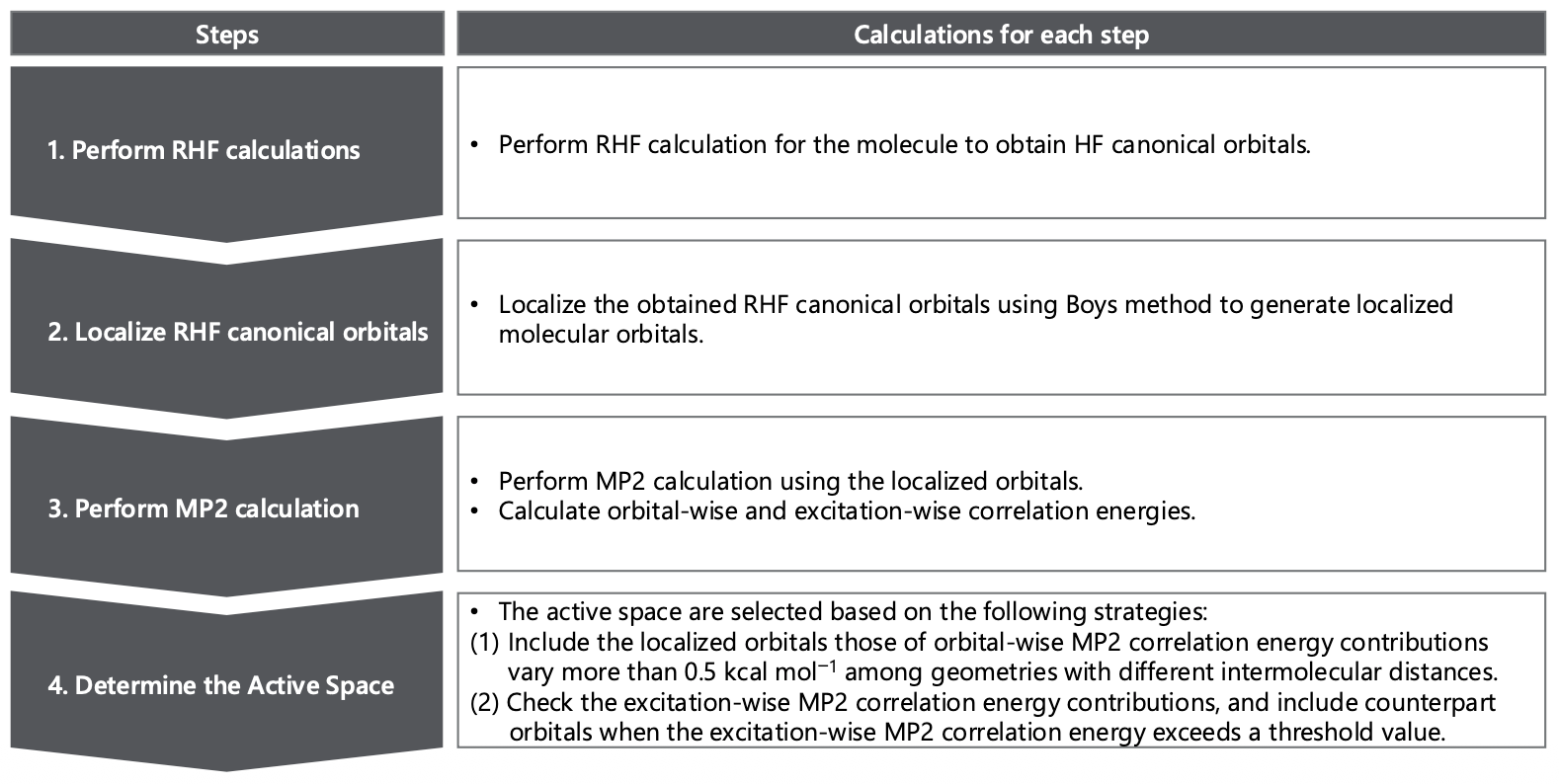}
    \caption{Steps for the MP2-based active space selection with Boys localized orbitals.}
    \label{fig:fig2}
\end{figure*}

The steps for the active orbital selection is given in Figure \ref{fig:fig2}. The steps begin with the Restricted Hartree--Fock (RHF) calculations, and then the RHF canonical orbitals are localized using Boys method. To ensure that the HF wave function remains invariant under orbital rotations, the occupied and unoccupied orbitals were localized separately. During the localization of the occupied orbitals, the molecular orbitals corresponding to 1s core orbital of the oxygen atoms were excluded from the localization procedure, to prevent mixing between core and valence orbitals. After that, the MP2 calculation using the localized orbitals is performed, and orbital-wise and excitation-wise correlation energies are calculated. Molecular orbitals to be included in the active space are selected based on the following strategies. (1) Include molecular orbitals those of the MP2-based orbital-wise correlation energy contributions vary more than 0.5 kcal mol$^{-1}$ among geometries with different intermolecular distances. (2) For each molecular orbital selected in step 1, check the excitation-wise MP2 correlation energy contributions, and include counterpart orbitals when the excitation-wise MP2 correlation energy exceeds a threshold value. The threshold value can be selected manually, by considering trade-off between accuracy and computational resource. By setting smaller threshold value, we can include more molecular orbitals in the active space and hence we can capture more correlation energy, but adding one molecular orbital results in addition of two qubits. 

\section{Computational conditions}
Geometry optimization of hydrogen-bonded water dimer was carried out at the PBE0~\cite{PBE0}/aug-cc-pVQZ~\cite{cc-pVQZ, aug-cc-pVQZ} level of theory, using PySCF software~\cite{PySCF}. No imaginary vibrational frequency was obtained at the optimized geometry. Cartesian coordinates of the water dimer are provided in Supporting Information. The intermolecular O$\cdots$O and H$\cdots$O distances at the hydrogen bonding site are calculated to be 1.929 and 2.889 \AA, respectively. To study potential energy curve for intermolecular hydrogen bonding, we generated six additional geometries by varying intermolecular H$\cdots$O distance relative to the equilibrium geometry by $\Delta r$ ($\Delta r$ = 100.00, 1.00, 0.75, 0.50, 0.25, and $-0.25$ \AA), while keeping all other geometrical parameters fixed. The geometry of $\Delta r = 100.00$ \AA\ is generated to check if the size consistency is maintained or not in the present computational conditions.  
For the supramolecular approach-based intermolecular interaction energy calculations, we took the water molecule monomers from the optimized geometry of the dimer, without performing geometry optimization for the monomer. 

The HF calculations and molecular orbital localizations were performed using PySCF~\cite{PySCF}, then fermionic Hamiltonian defined in eq (\ref{eq7}) is generated with the aid of openfermion-pyscf library~\cite{OpenFermion}. In this study, we used STO-3G basis set~\cite{STO-3G}. 
The MP2 calculations and orbital-wise and excitation-wise correlation energy analysis were carried out using our own Python3 code. 

For the QPE simulations, we adopted JWT~\cite{Jordan-1928} for the fermion--qubit encoding. The QPE quantum circuit simulation program is developed using Cirq library~\cite{Cirq}. We used six ancillary qubits and the evolution time length for the time evolution operator $U = e^{-i(H - E_\text{HF})t}$ is set to be $t = 128$. These QPE computational conditions were determined using information of the MP2 correlation energy, as discussed in Section IV-B. The second-order Trotter decomposition is applied to $U$ with the number of Trotter slices $M$ = 128, 256, and 512, to perform algorithmic error mitigation (AEM)~\cite{Endo-2019}. In the Trotter decomposition, we used a magnitude ordering for the term ordering~\cite{Tranter-2018}. In the QPE quantum circuit simulations, we adopted a strategy of sequential addition of ancillary qubits~\cite{Sugisaki-2024b}, to accelerate numerical simulations. 

\section{Results and Discussion}

\subsection{Orbital localization}
\begin{figure}
    \centering
    \includegraphics[width=\linewidth]
    {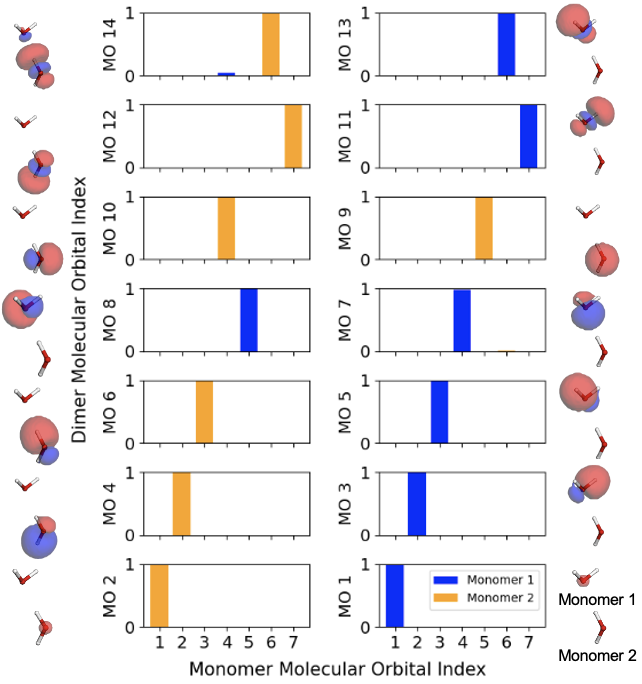}
    \caption{Overlap between the localized orbitals of monomer and dimer in the dimer equilibrium geometry and contour plots of the localized orbitals. MO means dimer molecular orbital, and the following number indicates the index.}
    \label{fig:fig3}
\end{figure}

At first, we focus on the preprocessing classical computation part of the QPE-based intermolecular interaction energy calculations, namely orbital localization. As we pointed out in the Theory section, orbital localization is essential to maintain size consistency in the QPE-CASCI. In order to reduce the computational cost by excluding unimportant molecular orbitals, the localized orbitals should satisfy the following conditions. (1) Molecular orbitals should be localized onto a monomer. (2) Almost the same localized orbitals should be obtained for the monomer and the dimer. (3) The shape of localized orbitals of the dimer should not be greatly dependent on the intermolecular distance. We first focus on the conditions (1) and (2). To assess the orbital localization method, we calculated the overlap between the localized orbitals of monomer and dimer in the dimer equilibrium geometry, $|\langle \psi_j^\text{monomer}|\psi_k^\text{dimer}\rangle|$. The results obtained using Boys localized orbitals in conjunction with contour plots of the localized orbitals are given in Figure \ref{fig:fig3}, and those using PM and ER localized orbitals are given in Figures \ref{fig:figS1} and \ref{fig:figS2}, respectively, in Supporting Information. Clearly, Boys method gave the molecular orbitals localized to the monomer. The confinement of the Boys localized orbitals to the monomer was also confirmed visually. Compared to Boys method, PM localized orbitals satisfy the second condition to a lesser extent, making it less suitable for the our objective. In contrast, ER method achieves a comparable level of localization to Boys method. This is a reasonable outcome, considering that PM method has a localization scheme based on the Mulliken atomic charges, whereas ER method, like Boys method, is based on the spatial distance.
The condition (3) is examined by calculating the overlap of the Boys localized orbitals of dimer in equilibrium and non-equilibrium geometries, $|\langle \psi_j^\text{dimer;Eq}|\psi_j^\text{dimer;non-Eq}\rangle|$. The orbital overlap using the HF canonical orbitals are also calculated for the reference. The results are plotted in Figures \ref{fig:figS3} and \ref{fig:figS4} in Supporting Information for HF canonical orbitals and Boys localized orbitals, respectively. Clearly, Boys localized orbitals have almost the same spatial distributions among all geometries being studied, while the shape of HF canonical orbital varies drastically by changing the intermolecular distance. In the present study, we used Boys localized orbitals for the QPE-based interaction energy calculations with the supramolecular approach. 

\subsection{Active orbital selection}
\begin{figure}
    \centering
    \includegraphics[width=\linewidth]
    {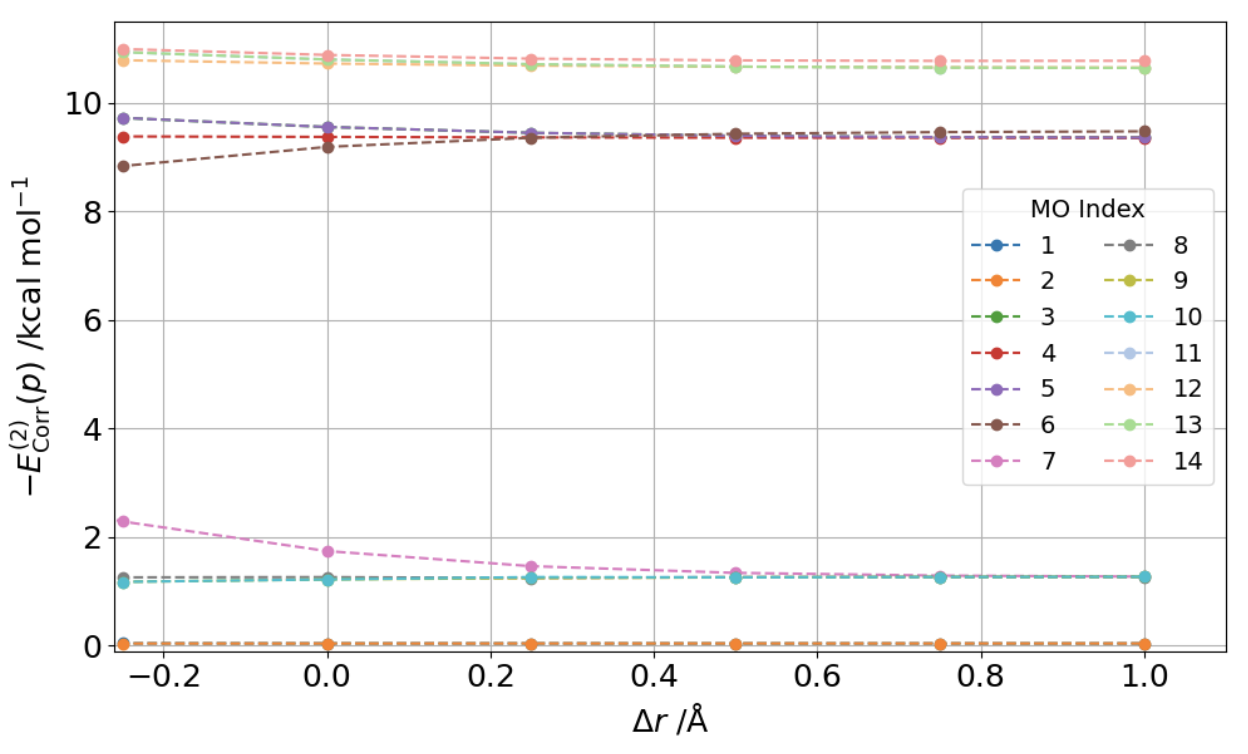}
    \caption{Orbital-wise MP2 correlation energy contribution. }
    \label{fig:fig4}
\end{figure}

Selection of appropriate active orbitals is a key step to obtain accurate intermolecular interaction energy as well as total energy. So far, active orbital selection based on chemist's intuition has been widely used, but applying this strategy sometimes face difficulties when the system is large and has a complicated electronic structure. Automated active orbital selection schemes based on density matrix renormalization group (DMRG)~\cite{Stein-2016} or machine learning~\cite{Jeong-2020} have been proposed, which often combined with orbital optimization process. In this work, we do not attempt to connect active orbital selection to orbital optimization, and we used a simpler approach based on the MP2 correlation energy contribution. To this end, we first computed orbital-wise contributions $E_\text{Corr}^{(2)}(p)$ by taking partial sum of the correlation energies. For example, for occupied orbitals, 
\begin{eqnarray}
    E_\text{Corr}^{(2)}(p) = \frac{1}{4}\sum_{jab} \frac{|g_{pjba} - g_{pjab}|^2}{\varepsilon_p + \varepsilon_j - \varepsilon_a - \varepsilon_b}.
    \label{eq10}
\end{eqnarray}
Note that $\sum_p E_\text{Corr}^{(2)}(p) \ne E_\text{Corr}^{(2)}$, because of duplicated counting of the correlation energy. $E_\text{Corr}^{(2)}(p)$ roughly estimates the loss of correlation energy when $p$-th molecular orbital is excluded from the active space. The results are summarized in Figure \ref{fig:fig4} and Table \ref{tab:table_from_fig4}. From these results, we can conclude that at least sixth and seventh localized molecular orbitals should be included in the active space, since $E_\text{Corr}^{(2)}(p)$ strongly depends on the intermolecular distance.  From Figure \ref{fig:fig3}, sixth and seventh localized orbitals are O--H $\sigma$ bonds along with the hydrogen bond, and they are strongly influenced by the intermolecular interaction. In the geometry of long intermolecular distance, molecular orbitals are roughly categorized to four groups, in the increasing order of the correlation energy contribution. (i) 1s core orbital of oxygen atoms (orbital index 1 and 2), (ii) lone pair orbitals of oxygen atoms (7--10), (iii) O--H $\sigma$ orbitals (3--6), and (iv) O--H $\sigma^*$ orbitals (11--14). It should be also noted that the behavior of $E_\text{Corr}^{(2)}(p)$ qualitatively agreed with that of the absolute value of the energy difference between the CASCI energy excluding the $p$-th molecular orbital from the active space and the full-CI energy (see Figure \ref{fig:figS5} in Supporting Information).

\begin{table}
\caption{\label{tab:table_from_fig4} Variation of the orbital-wise MP2 correlation energy among geometries with different intermolecualr distance.}
\begin{tabular*}{8.5cm}{@{\extracolsep{\fill}}cccc}
\hline
\multirow{2}{*}{MO index $p$} & $\max (-E_\text{Corr}^{(2)}(p))$ & $\min (-E_\text{Corr}^{(2)}(p))$ & difference$^{[a]}$ \\
 & /kcal mol$^{-1}$ & /kcal mol$^{-1}$ & /kcal mol$^{-1}$ \\
 \hline
3  &  9.725 &  9.362 & 0.363 \\
4  &  9.383 &  9.356 & 0.027 \\
5  &  9.726 &  9.362 & 0.364 \\
6  &  9.477 &  8.838 & 0.639 \\
7  &  2.288 &  1.273 & 1.015 \\
8  &  1.266 &  1.246 & 0.020 \\
9  &  1.270 &  1.177 & 0.093 \\
10 &  1.270 &  1.177 & 0.093 \\
11 & 10.936 & 10.645 & 0.291 \\
12 & 10.788 & 10.652 & 0.136 \\
13 & 10.937 & 10.645 & 0.292 \\
14 & 10.994 & 10.775 & 0.219 \\
\hline
\end{tabular*}
\raggedright
$^{[a]}$ $\max (-E_\text{Corr}^{(2)}(p)) - \min (-E_\text{Corr}^{(2)}(p))$.
\end{table}

Because both sixth and seventh localized orbitals are occupied ones, we have to include at least one virtual orbital in the active space to perform CASCI calculation. For this purpose, we calcualted the excitation-wise MP2 correlation energy contribution defined as
\begin{eqnarray}
    E_\text{Corr}^{(2)}(i,a) = \frac{1}{4} \sum_{jb} \frac{|g_{ijba} - g_{ijab}|^2}{\varepsilon_i + \varepsilon_j - \varepsilon_a - \varepsilon_b}.
    \label{eq11}
\end{eqnarray}

\begin{figure}
    \centering
    \includegraphics[width=\linewidth]
    {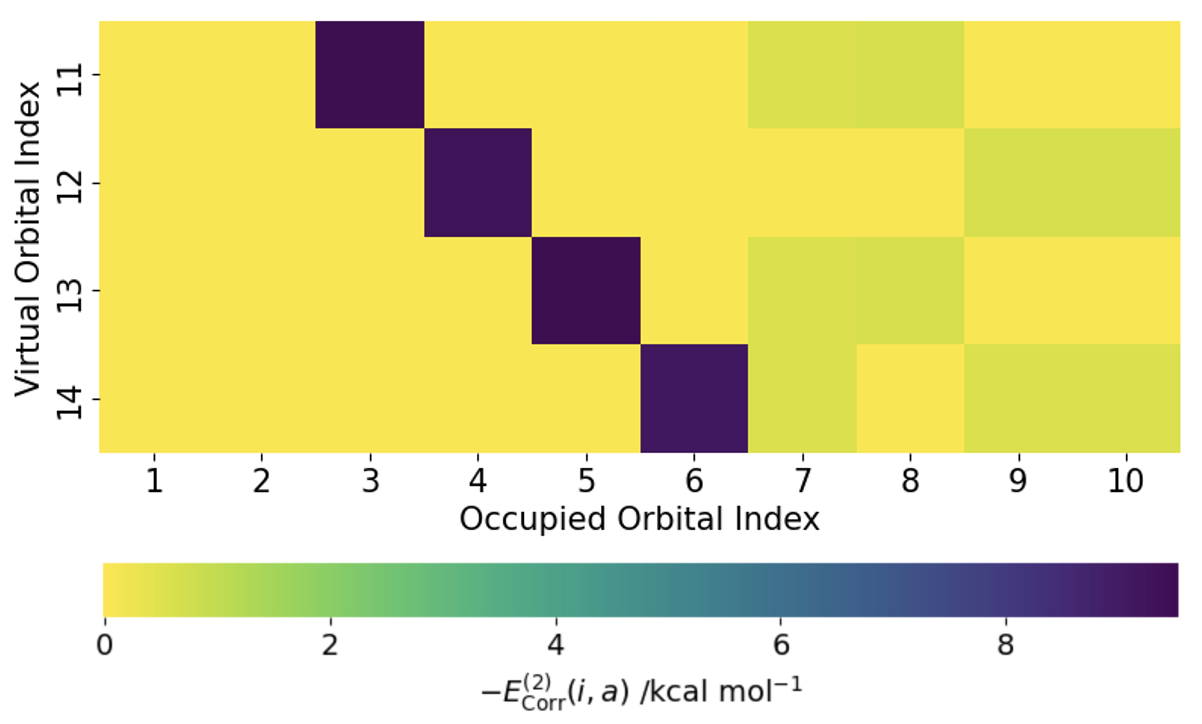}
    \caption{Excitation-wise MP2 correlation energy contribution.}
    \label{fig:fig5}
\end{figure}

The results of the $E_\text{Corr}^{(2)}(i,a)$ calculations are summarized in Figure \ref{fig:fig5}. Clearly, the most important excitations are $\sigma \rightarrow \sigma^*$ excitations of the O--H moieties (3 $\rightarrow$ 11, 4 $\rightarrow$ 12, 5 $\rightarrow$ 13, and 6 $\rightarrow$ 14). In contrast, excitations from lone pair orbitals are less significant. It should be noted that most of intermolecular (lone pair $\rightarrow \sigma^*$) excitations have negligible correlation energy contributions ($E_\text{Corr}^{(2)} \le 10^{-3}$ kcal mol$^{-1}$), but the excitation (7 $\rightarrow$ 14) have the correlation energy about three order of magnitude larger than others ($E_\text{Corr}^{(2)}(7, 14) = 0.58$ kcal mol$^{-1}$). This is consistent to the fact that the (7 $\rightarrow$ 14) excitation describes the electron donation caused by the hydrogen bond. From these analyses, we concluded that the smallest active space for the QPE-CASCI-based intermolecular interaction energy calculations is (4e, 3o) consisting of sixth, seventh, and 14th localized orbitals in Figure \ref{fig:fig3}, where ($N$e, $M$o) means that active space contains $N$ electrons and $M$ orbitals. The 11th and 13th localized orbitals should also be included if we set the threshold for the excitation-wise contribution being 0.5 kcal mol$^{-1}$. 

The QPE computational conditions are determined using information of the MP2 correlation energy. In our QPE setup, the evolution time $t$ can be set as $t \approx \frac{2\pi}{5\times |E_\text{Corr}^{(2)}|}$ in an atomic unit. For the (4e, 3o) active space, the MP2 correlation energy is $E_\text{Corr}^{(2)}(\text{4e, 3o}) = -0.011504$ hartree, and we set $t = 128$. If we adopt the phase $\phi'$ giving the maximum probability in the measurement in the QPE quantum circuit as the estimation of eigenphase, the rounding error in the energy estimation can be evaluated as follows.
\begin{eqnarray}
    \Delta \varepsilon \le \frac{1}{2} \times \frac{2\pi}{t2^{{N_a}}}
    \label{eq12}
\end{eqnarray}
Since we want to make $\Delta \varepsilon <$ 1 kcal mol$^{-1}$, we set $N_a = 6$ in the present study.

\subsection{Hamiltonian term truncation}

\begin{table*}
\caption{\label{tab:table1} Number of terms in qubit Hamiltonian and the energy error caused by Hamiltonian term truncations. }
\begin{tabular*}{\linewidth}{@{\extracolsep{\fill}}rccccrrrr}
\hline
\multirow{2}{*}{$\Delta r^\text{[a]}$} & \multicolumn{4}{c}{Number of qubit Hamiltonian terms} & \multicolumn{4}{c}{$\Delta E_{\text{truncated} - \text{untruncated}}/\rm{kcal\ mol^{-1}}$} \\
 & $\delta = 0.01$ & $\delta = 0.005$ & $\delta = 0.001$ & $\delta = 0.0005$ & $\delta = 0.01$ & $\delta = 0.005$ & $\delta = 0.001$ & $\delta = 0.0005$ \\
 \hline
 100.00 &  78 &  78 &  94 &  94 & $-$12.9646 & $-$12.9646 & $-$0.0003 & $-$0.0003 \\
   1.00 & 110 & 126 & 214 & 214 &     0.0654 &  $-$0.1888 &    0.0652 &    0.0652 \\
   0.75 & 122 & 162 & 214 & 230 &  $-$0.5693 &     0.1187 &    0.1914 &    0.1778 \\
   0.50 & 126 & 166 & 230 & 246 &  $-$1.2680 &     0.4819 &    0.4930 &    0.4935 \\
   0.25 & 166 & 198 & 246 & 294 &     1.2541 &     1.3818 &    1.2862 &    0.0020 \\
   0.00 & 150 & 230 & 294 & 326 &     3.0093 &     3.0785 &    0.0053 &    0.0000 \\
$-$0.25 & 194 & 230 & 326 & 342 &     6.4486 &     6.5174 &   $-$0.0002 &    0.0000 \\
%$-$0.25 &  78 &  82 & 182 & 182 &  $-$0.8902 &  $-$0.9493 &    0.0174 &    0.0000 \\
\hline
\end{tabular*}
\raggedright
$^{[a]}$The change in intermolecular H$\cdots$O distance from the equilibrium geometry.
\end{table*}

The qubit Hamiltonian $H_\text{qub}$ contains $\mathcal O(M^4)$ of Pauli strings, and the quantum circuit for the time evolution of single Pauli string $e^{-iP_jt}$ contains $\mathcal O(M)$ of CNOT gates when JWT is applied~\cite{Seeley-2012}. A randomization-based qDRIFT~\cite{Campbell-2019} is known to reduce the computational cost of the time evolution operation. In this work, we examined more simple approach by truncating Hamiltonian terms having small coefficient in absolute value. Namely, in the time evolution operation, we used the truncated Hamiltonian defined as 
\begin{eqnarray}
    \mathcal H = {\sum_{|h_{pq}| \ge \delta}}' h_{pq} a_p^\dagger a_q + \frac{1}{2}{\sum_{|g_{pqrs}| \ge \delta}}'g_{pqrs} a_p^\dagger a_q^\dagger a_s a_r
    \label{eq13}
\end{eqnarray}
where $\delta$ is the threshold value. It is clear that $[\mathcal H, H] \ne 0$ and therefore energy eigenvalue and eigenfunction of $\mathcal H$ is different from those of $H$. In order to obtain intermolecular interaction energy using QPE with a chemical precision, the difference of eigenvalues of $\mathcal H$ and $H$ should be smaller than 1 kcal mol$^{-1}$. In this work, we performed Hamiltonian matrix diagonalization (CASCI calculation) with four different threshold values, $\delta$ = 0.01, 0.005, 0.001, and 0.0005, and checked the number of remaining Pauli strings and the error in energy caused by the Hamiltonian term truncations. The results are summarized in Table \ref{tab:table1}. The number of terms of the untruncated Hamiltonian is 342. We observed that the effect of Hamiltonian term truncation strongly depends on the molecular geometry, and using the same threshold value $\delta$ for all geometries is problematic unless $\delta$ is as small as 0.0005. Importantly, the error in energy does not vary monotonically with $\delta$, and using tighter threshold sometimes results in larger error. In the present study, we started with very small active space containing only minimum necessary active orbitals, hence Hamiltonian term truncation is not a good option to begin with. In the QPE simulations discussed below, we used untruncated Hamiltonian for the time evolution operator.  

\subsection{QPE simulations}

\begin{figure}
    \centering
    \includegraphics[width=\linewidth]{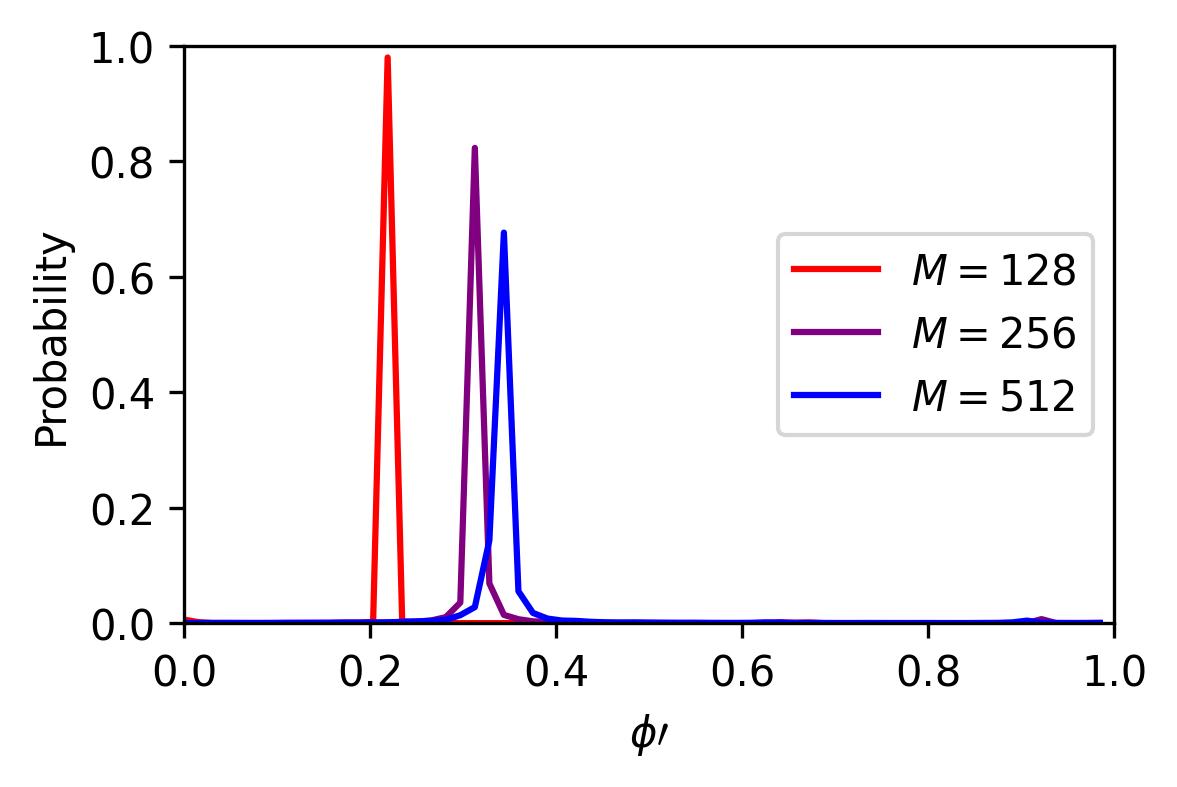}
    \caption{Phase value vs probability plot obtained from QPE simulation of the water dimer in its equilibrium geometry. Number of ancilla qubit is 6, $t = 128$, and the number of Trotter slice is 128 (red), 256 (purple), and 512 (blue).}
    \label{fig:fig6}
\end{figure}

\begin{table}
\caption{\label{tab:table2} Error of the QPE energy of water dimer after AEM with respect to the CASCI one.}
\begin{tabular*}{8.5cm}{@{\extracolsep{\fill}}rrr}
\hline
\multirow{2}{*}{$\Delta r^\text{[a]}$} & $\Delta E(\text{MaxProb})$ & $\Delta E(\text{WgtAve})$ \\ 
 & /kcal mol$^{-1}$ & /kcal mol$^{-1}$ \\
 \hline
 100.00 &    0.0849 & $-$0.0390  \\
   1.00 &    0.1794 &    0.0725 \\
   0.75 & $-$0.0930 & $-$0.0695 \\ 
   0.50 & $-$0.1568 & $-$0.0586 \\
   0.25 &    0.0294 &    0.0098 \\
   0.00 & $-$0.0823 & $-$0.0586 \\
$-$0.25 & $-$0.1540 & $-$0.1048 \\
%$-$0.25 &    0.1111 &    0.0971 \\
\hline
\end{tabular*}
\raggedright
$^{[a]}$The change in intermolecular H$\cdots$O distance from the equilibrium geometry.
\end{table}

Figure \ref{fig:fig6} represents phase value $\phi'$ vs probability plot obtained from the QPE simulation of the water molecule dimer in its equilibrium geometry with three different Trotter slices. Since we used $t = 128$ for the time evolution operator $U$ and $N_a = 6$ of ancillary qubits, the rounding error of the phase value is expected to be $\Delta \varepsilon \le 0.2407$ kcal mol$^{-1}$ when the grid point giving the maximum probability is used as the estimate of $\phi$. More accurate estimation of $E'$ from the probability plot can be obtained by taking weighted average of several grid points around the peak. In this work, we examined two strategies for the $\phi'$ estimation, namely using the grid value of maximum probability (MaxProb) and weighted average of three grid points around the peak (WgtAve). Once the energy estimation is obtained for given number of Trotter slice, then AEM is applied to estimate the Trotter error-free energy. It is known that the error in the second-order Trotter decomposition scales as $\mathcal{O}((t/M)^2)$. In the AEM, the $1/M$ vs. $E'$ plot is fitted by a quadratic function $E' = a(1/M)^2 + b$, and used $b$ as a Trotter-free value. Detail of the AEM procedure is given in Supporting Information. The error of the QPE energy after AEM of the water dimer with different geometries with respect to the CASCI energy is summarized in Table \ref{tab:table2}. The error of the QPE energy is smaller when weighted average strategy is adopted, indicating that classical postprocessing to compute $E'$ from the probability plot is important to obtain accurate energy. In the following discussion on the interaction energies, we use $E$(AEM;WgtAve) as the QPE energy. 
The electron correlation energy in the water dimer in its equilibrium geometry calculated using QPE with AEM is $-0.017092$ hartree, which is about 1.5 times larger than that captured in the MP2 method with the same active space. The CASCI correlation energy is $-0.016998$ hartree, and therefore percentage error of the QPE correlation energy is only about 0.55\%. Total energy of the dimer in the equilibrium geometry is calculated to be $-149.952230$ hartree.

To compute the interaction energy using supramolecular approach, we need energies of the monomers. Note that in the (4e, 3o) active space for the dimer, one of the water molecules (monomer 1) has only one occupied orbital in the active space, and the CASCI is equivalent to the RHF for monomer 1. The RHF energy of monomer 1 is $-74.962844$ hartree. For monomer 2, we carried out the QPE simulations, with the same computational conditions as the dimer calculations. After adopting AEM, the QPE energy of the monomer 2 is calculated as $-74.981175$ hartree, and the error of the QPE energy from the CASCI value is $-0.0389$ kcal mol$^{-1}$. As a result, the interaction energy of the hydrogen bonding in the water molecule dimer is computed as $-5.1333$ kcal mol$^{-1}$. This value should be compared with the CASCI value, $-5.1530$ kcal mol$^{-1}$. Note that the interaction energy at the RHF level is $E_\text{int}$(RHF) = $-5.7017$ kcal mol$^{-1}$. Although the effect of electron correlation on the interaction energy is small in hydrogen bonding, our QPE simulation succeed to capture such a small electron correlation effect.
The intermolecular interaction energy calculated at the full-CI level is $E_\text{int}\text{(full-CI)} = -5.6231$ kcal mol$^{-1}$. Consequently, the interaction energy at the RHF level is closer to the full-CI value than that at the CASCI level. It is well known that hydrogen bonding is primarily governed by electrostatic interactions, so the effect of electron correlation is not significant~\cite{Rezac-2011, Rezac-2014} However, the situation may differ for other types of intermolecular interactions.

As discussed above, we used localized orbitals not only to select molecular orbitals relevant to intermolecular interactions, but also to ensure size consistency in QPE calculations. To assess the size consistency of the present QPE simulations, we compared the energy of the dimer at $\Delta r$ = 100 \AA\ with the sum of the energies of monomer 1 and monomer 2. $E(\Delta r = 100$ \AA) is calculated to be $-149.943956$ hartree, while ($E$(monomer 1) + $E$(monomer 2)) = $-149.944018$ hartree. The difference between these values is 0.039 kcal mol$^{-1}$, which is sufficiently small for the discussion of intermolecular interaction energies. This small deviation may originate from the postprocessing method used for eigenphase readout, or slight changes in the shapes of the active orbitals. The interaction energy calculated as the difference between the dimer energies at the equilibrium and at $\Delta r$ = 100 \AA\ is $-5.1919$ kcal mol$^{-1}$.

The most cost-demanding part of the QPE quantum circuit is the controlled-time evolution operation. The number of two-qubit gates and depth of the single Trotter slice of the controlled-$U$ gate of the water dimer is 1494 and 1704, respectively. From this, the depth and the number of two-qubit gates of the entire QPE quantum circuit of the water dimer with $N_a = 6$, $t = 128$, and $M = 128$ is estimated to be $1.20 \times 10^7$ and $1.37 \times 10^7$, respectively. This undoubtedly exceeds the capability of current quantum hardware, and classical optimizations of quantum circuit based on tensor network-based approaches~\cite{Kanno-2025} or using publicly available software such as qiskit~\cite{qiskit}, Qmod~\cite{Qmod}, PyZX~\cite{PyZX}, and pytket~\cite{PyTKET}, in conjunction with pipeline optimization methods~\cite{Kharkov-2022, Baskaran-2023} are necessary to reduce the number of two-qubit gates and depth. Using the same QPE conditions, we conducted a preliminary investigation of quantum circuit compression with Classiq SDK and Qmod~\cite{Qmod}, achieving reductions of over 99\% in both circuit depth and the number of two-qubit gates More detailed analysis of the quantum circuit compression is underway. 

\section{Conclusion}
In this study, we demonstrated the resource-efficient QPE-CASCI simulations of hydrogen-bonded water dimer and monomers to calculate intermolecular interaction energy in terms of a supramolecular approach, in conjunction with the MP2-based active orbital selections with Boys localized orbitals and algorithmic error mitigation. By applying the proposed methods, we have successfully reduced the number of qubits required to calculate the potential energy curve of hydrogen bonding quantitatively to 12 (six qubits for wave function storage and six qubits for ancilla). Further reduction of the computational resources is of course possible, for instance, by applying SCBKT instead of JWT to reduce two qubits for wave function storage~\cite{Bravyi-2017}. The correlation energies calculated using QPE agreed with the CASCI values with less than 1\% of error. The interaction energy calculated using QPE with supramolecular approach is $E_\text{int} = -5.1333$ kcal mol$^{-1}$, which agreed with the CASCI value of the same active space ($E_\text{int}\text{(CASCI)} = -5.1530$ kcal mol$^{-1}$), exemplifying effectiveness of the proposed approach. The importance of classical optimization of the QPE quantum circuit is also demonstrated. 

The present study is based on the textbook-implementation of QPE, which typically requires about ten ancillary qubits to read out the eigenphase in chemical precision. This prevents application of the proposed workflow to larger systems. It is known that the number of ancillary qubits can be reduced to one, by adopting other algorithms in QPE family, such as iterative QPE~\cite{Dobsicek-2007}, Bayesian QPE~\cite{Wiebe-2016}, and Hadamard test-based implementations~\cite{Ni-2023, Ding-2023}. Exploring these methods with supramolecular approach and applications to other types of intermolecular interactions are in progress. It should be also noted that even with classical optimization of the QPE quantum circuit is performed, the circuit depth and number of quantum gate can easily exceed the capability of NISQ devices when a larger active space is employed. In this context, estimating quantum computational resources in fault-tolerant quantum computing settings using Clifford + T gate sets~\cite{TFermion, Nelson-2024} becomes an important issue for assessing QPE-based interaction energy calculations. This topic will be discussed in the forthcoming paper. 

\section{Acknowledgments}
This work was partially supported by the Center of Innovation for Sustainable Quantum AI (Grant No. JPMJPF2221) from Japan Science and Technology Agency (JST), Japan. K.S. acknowledges the partial support from KAKENHI Transformative Research Area B (23H03819) and Scientific Research C (21K03407) from Japan Society for the Promotion of Science (JSPS), Japan. We acknowledge Dr. Shahaf Asban and Dr. Genki Okano at Classiq Technologies Ltd. for their technical support related to Classiq SDK and Qmod.

\section{Conflicts of interest}
The authors declare no conflict of interest. The use of Classiq SDK and Qmod in this study was based solely on publicly available or licensed access, and there was no financial or institutional relationship that influenced the outcomes of this research.

\bibliography{ref}
\bibliographystyle{unsrt}

\clearpage
\section{Supporting Information}

\renewcommand{\thepage}{S\arabic{page}}
\setcounter{page}{1}
\renewcommand{\thefigure}{S\arabic{figure}}
\setcounter{figure}{0}
\renewcommand{\thetable}{S\arabic{table}}
\setcounter{table}{0}
\vspace{-10pt}
\subsection{Cartesian coordinates}

\begin{table}[H]
\centering
    \vspace{-28pt}
    \caption{Cartesian coordinates of the PBE0/aug-cc-pVQZ optimized geometry of water molecule dimer, in units of {\r A}.}
    \begin{tabular*}{\linewidth}{@{\extracolsep{\fill}}crrr}
    \hline
    Atom & \multicolumn{1}{c}{X} & \multicolumn{1}{c}{Y} & \multicolumn{1}{c}{Z} \\
    \hline
    O & $-$2.54176 &    0.57839 & $-$0.53932 \\
    H & $-$1.81161 &    1.15639 & $-$0.76640 \\
    H & $-$2.99305 &    1.01380 &    0.18562 \\
    O & $-$1.47125 & $-$1.92083 &    0.43805 \\
    H & $-$1.80326 & $-$2.65320 & $-$0.08029 \\
    H & $-$1.87142 & $-$1.13254 &    0.04768 \\
    \hline
    \end{tabular*}
    \vspace{-5pt}
    \label{tab:table_s1}
\end{table}

\subsection{Localized orbitals}

\begin{figure}[h!]
    \centering
    \vspace{-20pt}
    \includegraphics[scale=0.65]
    {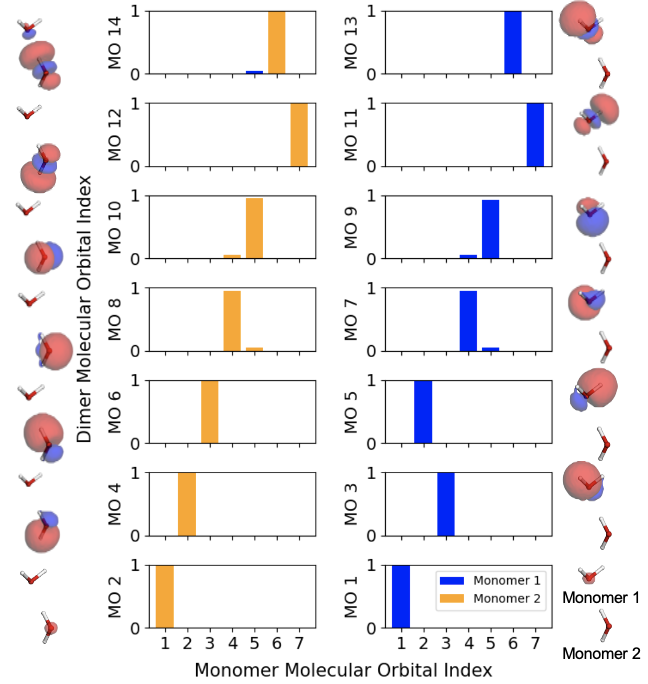}
    \caption{Overlap between the PM localized orbitals of monomer and dimer in the dimer equilibrium geometry and contour plots of the localized orbitals.}
    \label{fig:figS1}
    \vspace{-20pt}
\end{figure}

\begin{figure}[H]
    \centering
    \includegraphics[scale=0.65]
    {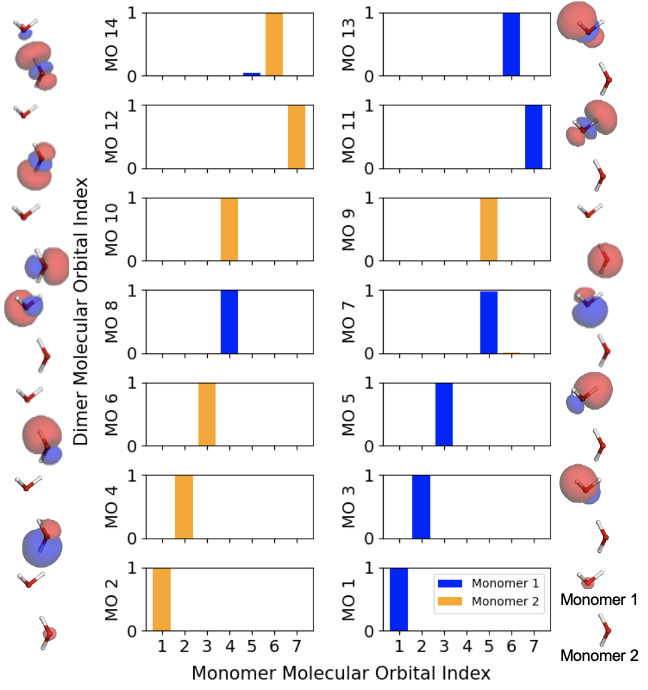}
    \caption{Overlap between the ER localized orbitals of monomer and dimer in the dimer equilibrium geometry and contour plots of the localized orbitals.}
    \vspace{-20pt}
    \label{fig:figS2}
\end{figure}

\begin{figure}[h!]
    \centering
    \includegraphics[width=0.9\linewidth]
    {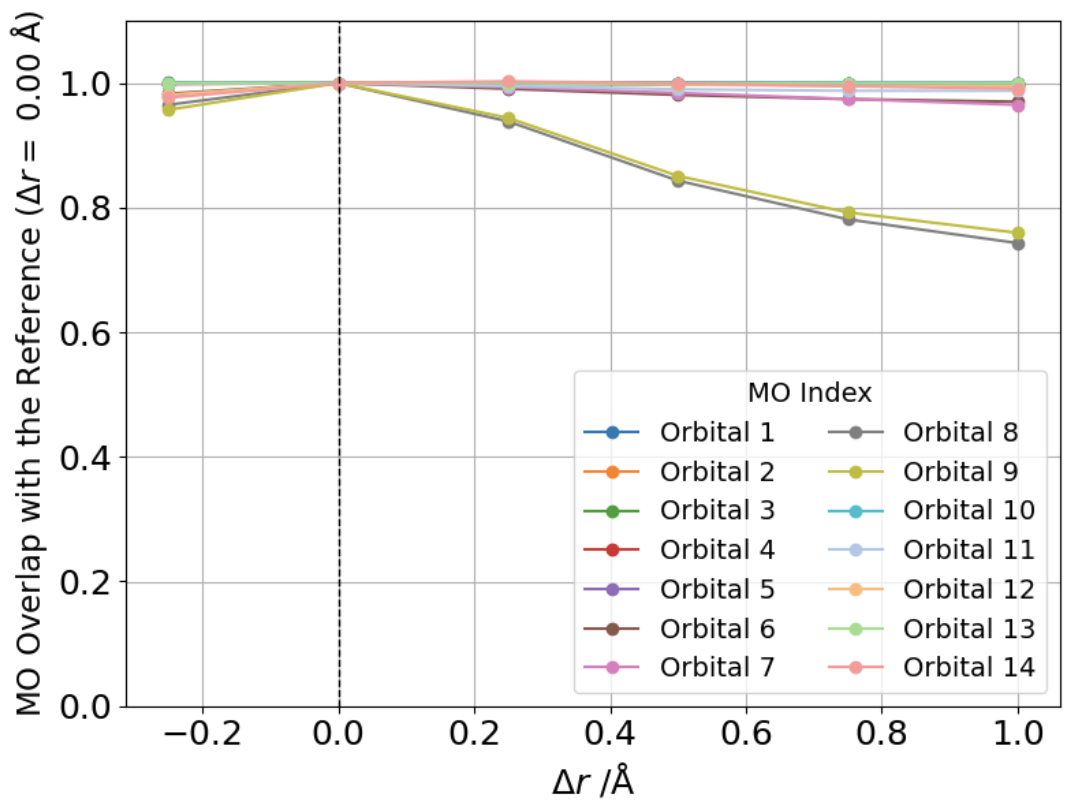}
    \caption{Overlap between the HF canonical orbitals of dimer in equilibrium ($\Delta r=$0.00 $\mathrm{\AA}$) and non-equilibrium geometries.}
    \label{fig:figS3}
    \vspace{-20pt}
\end{figure}

\begin{figure}[h!]
    \centering
    \includegraphics[width=0.9\linewidth]
    {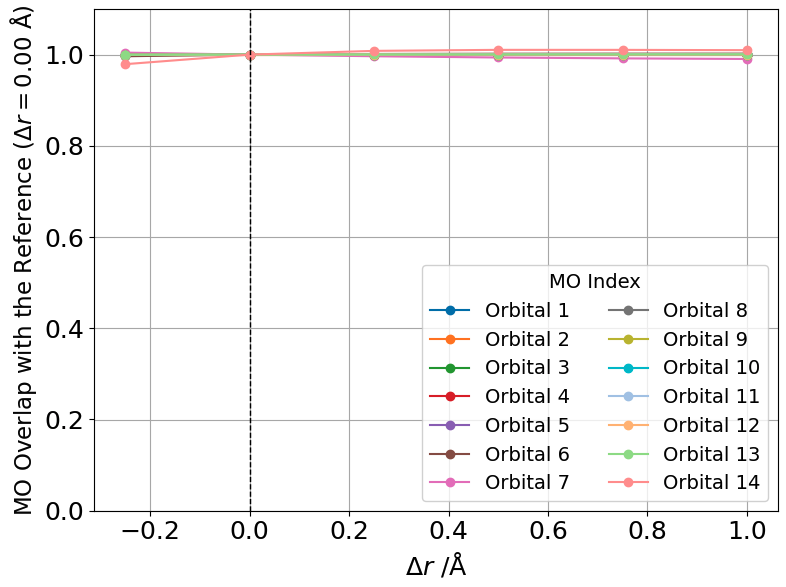}
    \caption{Overlap between the Boys localized orbitals of dimer in equilibrium ($\Delta r=$0.00 $\mathrm{\AA}$) and non-equilibrium geometries.}
    \label{fig:figS4}
    \vspace{-30pt}
\end{figure}

\subsection{Absolute value of the energy difference between the CASCI energy and the full-CI energy}

\begin{figure}[h!]
    \centering
    \vspace{-20pt}
    \includegraphics[width=0.9\linewidth]
    {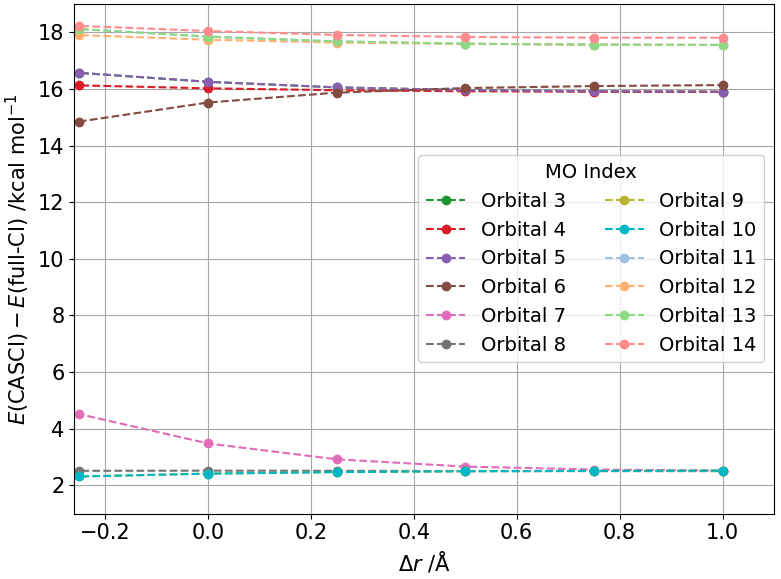}
    \caption{Absolute value of the energy difference between the CASCI energy excluding the $p$-th molecular orbital from the active space and the full-CI energy. For all CASCI calculations, the first and second molecular orbitals were frozen.}
    \vspace{-20pt}
    \label{fig:figS5}
\end{figure}

\newpage
\subsection{Algorithmic error mitigation}
\begin{figure}[H]
    \centering
    \includegraphics[width=\linewidth]{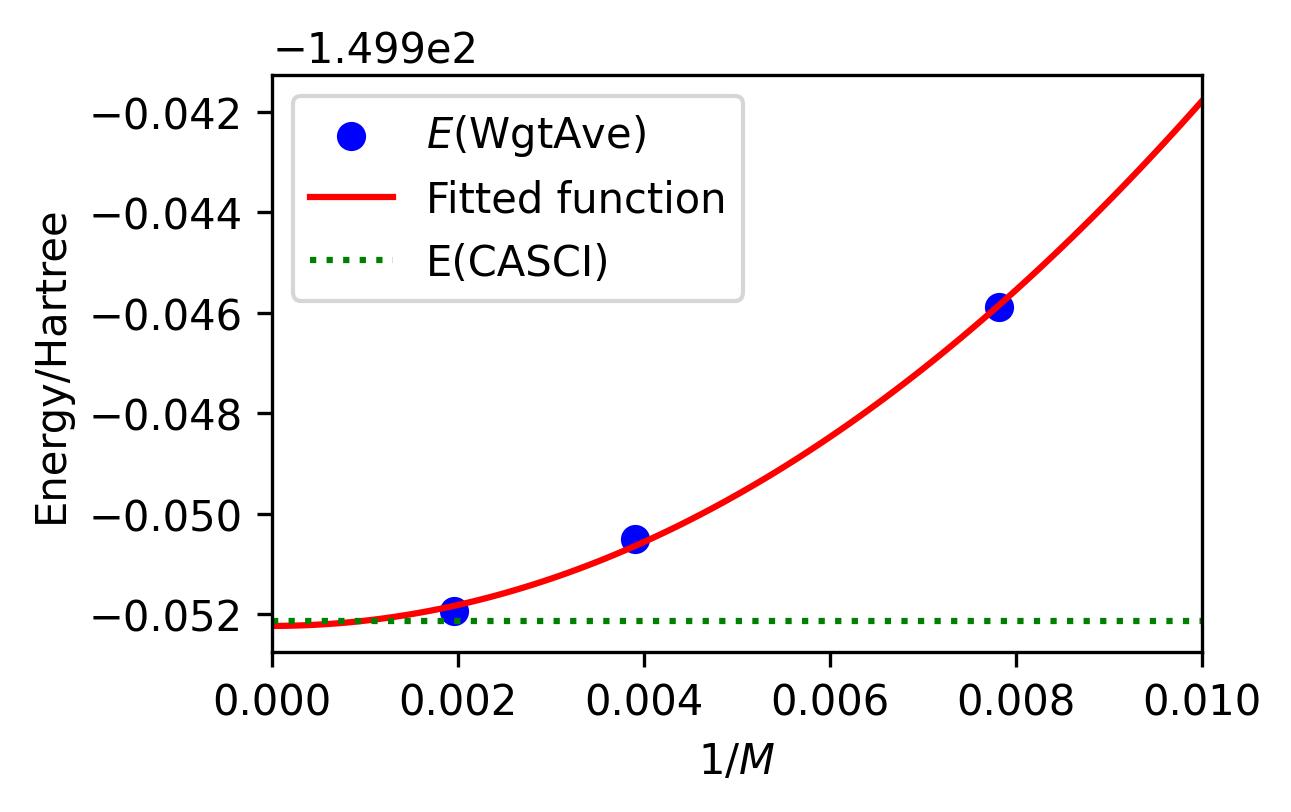}
    \caption{AEM result of the QPE energy of the water dimer in its equilibrium geometry.}
    \label{fig:figS6}
\end{figure}

\vspace{-20pt}
Algorithmic error mitigation (AEM) is a technique to mitigate algorithmic-origin errors such as a Trotter decomposition error. As discussed in the Results and Discussion section, an error of the second-order Trotter decomposition is expected to scale as $\mathcal{O}((t/M)^2)$. From this, we can extrapolate Trotter-free QPE energy by plotting the $1/M$ vs $E'$ and fitting the points by a quadratic function $E' = a(1/M)^2 + b$. The fitting is carried out using curve\_fit function in SciPy~\cite{2020SciPy-NMeth}. An example of the AEM in the case of the water dimer in its equilibrium geometry is given as Figure~\ref{fig:figS6}. The fitted parameters are $a = 104.513583$ and $b = -149.952230$, with the variances of the parameter estimate being $1.54 \times 10^{1}$ and $2.04 \times 10^{-8}$ for $a$ and $b$, respectively. The intercept of the fitting function $b$ corresponds to the estimated Trotter-free QPE energy. In the equilibrium geometry of the water dimer, the CASCI energy is calculated to be $E\mathrm{(CASCI)} = -149.952137$ hartree, and the difference between QPE energy after AEM and CASCI is $-0.0586$ kcal mol$^{-1}$. AEM for other geometries are carried out in the same way.

\end{document}